\def\year{2022}\relax
\documentclass[letterpaper]{article} % DO NOT CHANGE THIS
\usepackage{aaai21}  % DO NOT CHANGE THIS
\usepackage{times}  % DO NOT CHANGE THIS
\usepackage{helvet} % DO NOT CHANGE THIS
\usepackage{courier}  % DO NOT CHANGE THIS
\usepackage[hyphens]{url}  % DO NOT CHANGE THIS
\usepackage{graphicx} % DO NOT CHANGE THIS
\usepackage{csquotes}
\usepackage{amssymb}
\usepackage{rotating}
\usepackage{amsmath}
\usepackage{natbib}  % DO NOT CHANGE THIS AND DO NOT ADD ANY OPTIONS TO IT
\usepackage{caption} % DO NOT CHANGE THIS AND DO NOT ADD ANY OPTIONS TO IT
\usepackage{array}
\usepackage{hhline}
\usepackage{setspace} 
\usepackage{makecell}
\usepackage[flushleft]{threeparttable}
\usepackage{booktabs,caption}
\usepackage[toc,page]{appendix}
\usepackage{subcaption}
\usepackage{rotating}
\usepackage{multirow}
\usepackage[T1]{fontenc}

\title{Understanding Political Polarization via Jointly Modeling Users, Connections and Multimodal Contents on Heterogeneous Graphs}
\author {
    Hanjia Lyu, \textsuperscript{\rm 1}
    Jiebo Luo \textsuperscript{\rm 1,2}\\
}
\affiliations{
    \textsuperscript{\rm 1} University of Rochester\\
    \textsuperscript{\rm 2} Meta AI\\
    hlyu5@ur.rochester.edu, jluo@cs.rochester.edu\\}
\begin{document}


\maketitle

\begin{abstract}
Understanding political polarization on social platforms is important as public opinions may become increasingly extreme when they are circulated in homogeneous communities, thus potentially causing damage in the real world. Automatically detecting the political ideology of social media users can help better understand political polarization. However, it is challenging due to the scarcity of ideology labels, complexity of multimodal contents, and cost of time-consuming data collection process. Most previous frameworks either focus on unimodal content or do not scale up well. In this study, we adopt a heterogeneous graph neural network to jointly model user characteristics, multimodal post contents as well as user-item relations in a bipartite graph to learn a comprehensive and effective user embedding without requiring ideology labels. We apply our framework to online discussions about economy and public health topics. The learned embeddings are then used to detect political ideology and understand political polarization. Our framework outperforms the unimodal, early/late fusion baselines, and homogeneous GNN frameworks by a margin of at least \textbf{9\% absolute gain} in the area under the receiver operating characteristic on two social media datasets. More importantly, our work does \textbf{\textit{not}} require a time-consuming data collection process, which allows faster detection and in turn allows the policy makers to conduct analysis and design policies in time to respond to crises. We also show that our framework learns meaningful user embeddings and can help better understand political polarization. Notable differences in user descriptions, topics, images, and levels of retweet/quote activities are observed. Our framework for decoding user-content interaction shows wide applicability in understanding political polarization. Furthermore, it can be extended to user-item bipartite information networks for other applications such as content and product recommendation.
\end{abstract}

\section{Introduction}
A 2016 Pew Research study~\cite{gottfried2016news} found that of all U.S. adults, 67\% use social media platforms with 44\% using the platforms to discover news. Social media is found to shape political discourse in the U.S. and the whole world~\cite{o2010tweets,tumasjan2010predicting,conover2011political, weld2021political}. The extent to which the opinions on an issue are opposed is termed ``political polarization''~\cite{dimaggio1996have}. The formation of such political polarization is not necessarily a serious problem, but the concern is that the opinions may become increasingly polarized when they are shared, viewed and discussed in a homogeneous community~\cite{conover2011political}. For example, on March 16, 2020, the former U.S. President Donald Trump posted a tweet calling COVID-19 ``Chinese Virus''.\footnote{\url{https://twitter.com/realdonaldtrump/status/1239685852093169664} [Accessed 2020-03-16]} After that, this term was frequently used among the right-leaning Twitter users, and in most cases, they were carried with negative sentiment against Chinese and Asian people~\cite{lyu2020sense, chen2021fine}. More importantly, the harm caused by biased and false news has a substantial negative impact across the entire society~\cite{weld2021political}. Fake or extremely biased news regarding COVID-19 vaccines is found to be negatively correlated with the state-level COVID-19 vaccination rate~\cite{lyu2022misinformation}. The growing political divide in the U.S. also extends to the \#StopAsianHate movement with Trump's followers being less supportive to the Asian-American victims of the racially motivated hate crimes~\cite{lyu2021understanding}. Therefore, understanding political polarization on social platforms is important.

To better understand political polarization on social platforms, we also need to measure political ideology which refers to the political stance of people~\cite{xiao2020timme}. It is noteworthy that political ideology is a complicated concept and may often be used at multiple levels~\cite{feldman2013political}. Similar to previous studies~\cite{conover2011predicting,chen2017opinion,xiao2020timme}, we frame the stance detection problem as ideology detection. By identifying people's ideology, we would be able to investigate questions such as what opinions they hold or what characteristics they share. There are two major challenges in estimating political ideology in real-world applications. First, unlike some politicians who explicitly show their ideology, the political ideology of most ordinary citizens remains unknown or hidden. Although social platforms such as Twitter allow users to describe themselves in the profile description section, not all users indicate whether they are left- or right-leaning explicitly. Their views are critical when it comes to predicting the outcomes of political events such as elections~\cite{xiao2020timme,wang2017follow}. It is expensive to obtain human annotated labels for political ideology. Second, each user is often associated with multiple sources of information (e.g., user characteristics, multimodal post contents, relations with other users and items), however, most existing works of ideology detection on social platforms only focus on one of them and have drawbacks including (a) limited robustness with potentially noisy unimodal information on social media~\cite{jin2017multimodal} and (b) difficulty in providing real-time detection due to a prohibitively costly data collection process~\cite{xiao2020timme} that may cause a delay in the decision-making of the policy makers. Therefore, a viable framework should be able to detect political ideology in real time while maintaining high performance. The first challenge also limits our understanding of online political polarization as most previous studies are based on unimodal information.

In this study, we address the above challenges by introducing an efficient framework based on bipartite heterogeneous graph neural networks (GNN) to learn user embeddings \textbf{without requiring political ideology labels}. The learned embeddings are then used to detect the political ideology of social media users and understand online political polarization. We apply our framework to two large Twitter datasets that record the user-item interaction networks about two major current topics. Our study consists of three parts:

\textbf{Joint modeling:} We learn user embeddings by jointly modeling user characteristics, multimodal post contents, as well as user-item relations using a bipartite heterogeneous graph neural architecture in a weakly-supervised manner. A social-platform based negative sampling strategy is proposed to find important negative samples. We apply our framework to user-tweet bipartite graphs on Twitter. However, it can be applied to general user-item bipartite information networks such as user-microvideo graphs on TikTok and user-post graphs on Facebook.

\textbf{Detecting political ideology:} We use the user embeddings learned by our framework for an ideology detection task. We show that using the learned representations can outperform the unimodal, early/late fusion baselines, and homogeneous GNN frameworks by a margin of at least \textbf{9\% absolute gain} in the area under the receiver operating characteristic (AUROC) in two social media datasets, and more importantly, our framework does \textbf{\textit{not}} require a time-consuming data collection process as the previous work~\cite{xiao2020timme} does, which makes it possible to learn the ideology of the users who participate in political discussions and understand their opinions faster and in turn, to design policies in time to respond to major events.

\textbf{Understanding political polarization:} Our framework learns meaningful user embeddings and can help better understand online political polarization. We conduct a fine-grained analysis and find there is clear segregation between left- and right-leaning users. Notable differences in user descriptions, topics, images, and levels of retweet/quote activities are observed both within the same ideology group and across the different ideology groups.

The remainder of the paper is organized as follows. Related work is summarized in Section 2. In Section 3, we describe two datasets and the design of our framework. Empirical evidence is provided to validate our framework in estimating political ideology of social media users in Section 4. We also conduct a characterization study to provide insights into online political polarization. Finally, we discuss future work and conclude the paper in Section 5.

\section{Related Work}

\subsection{Political ideology detection}
Social platforms have rich multimodal information. Most studies rely on textual information to detect the political ideology of social media users~\cite{conover2011predicting,chen2017opinion,preoctiuc2017beyond}. However, the unimodal information is usually noisy on social media~\cite{jin2017multimodal} and can hinder the detection performance. Inspired by the findings of previous work that multimodal information is helpful when the unimodal signal is noisy~\cite{gurban2008dynamic, yuhas1989integration, wang2017polarized}, our framework leverages the multimodal post contents to tackle the ideology detection problem. Recently, although still using unimodal information, \citet{xiao2020timme} show promising performance in predicting people's ideology by modeling the relations (e.g., follow) among users in a heterogeneous network. Our framework also builds upon a heterogeneous network design. However, ours differ from \citet{xiao2020timme} fundamentally as our framework (1) takes in multimodal information, (2) considers two types of nodes (i.e., user nodes and tweet nodes) in the network, (3) only focuses on the user-item relations which does not require a time-consuming data collection process, and (4) is trained without political ideology labels, while their approach (1) takes in unimodal information, (2) considers only one type of node (i.e., user node), (3) models user-item relations and user-user relations while collecting user-user relations requires an extra data collection process which is time-consuming, and (4) is trained with political ideology labels.

\subsection{Political polarization understanding}
Most prior works on learning political polarization on social platforms focus on user behavior and unimodal contents~\cite{hosseinmardi2021examining, borge2015content,gruzd2014investigating, guarino2020characterizing, conover2011political,waller2021quantifying,urman2020context}. \citet{borge2015content} analyze the Egyptian political polarization on Twitter by a set of manually labeled hashtags and the retweet network constructed by hand-labeled seed users. \citet{hosseinmardi2021examining} identify several distinct communities of news consumers, including ``far-right'' and ``anti-woke'' by examining the consumption of radical content on YouTube. \citet{conover2011political} investigate how social media shape the networked public sphere and facilitate communication between communities with different political orientations. In particular, they use the 2010 U.S. congressional midterm elections as a case study and focus on the retweet and mention networks. \citet{waller2021quantifying} quantify the online political polarization using community-level embeddings. In contrast, our individual-level embeddings learned by jointly modeling user characteristics, multimodal contents, and user-item relations allow us to conduct a more fine-grained analysis.

\subsection{Graph neural networks}
\begin{figure}[htbp]
    \centering
    \includegraphics[width =\linewidth]{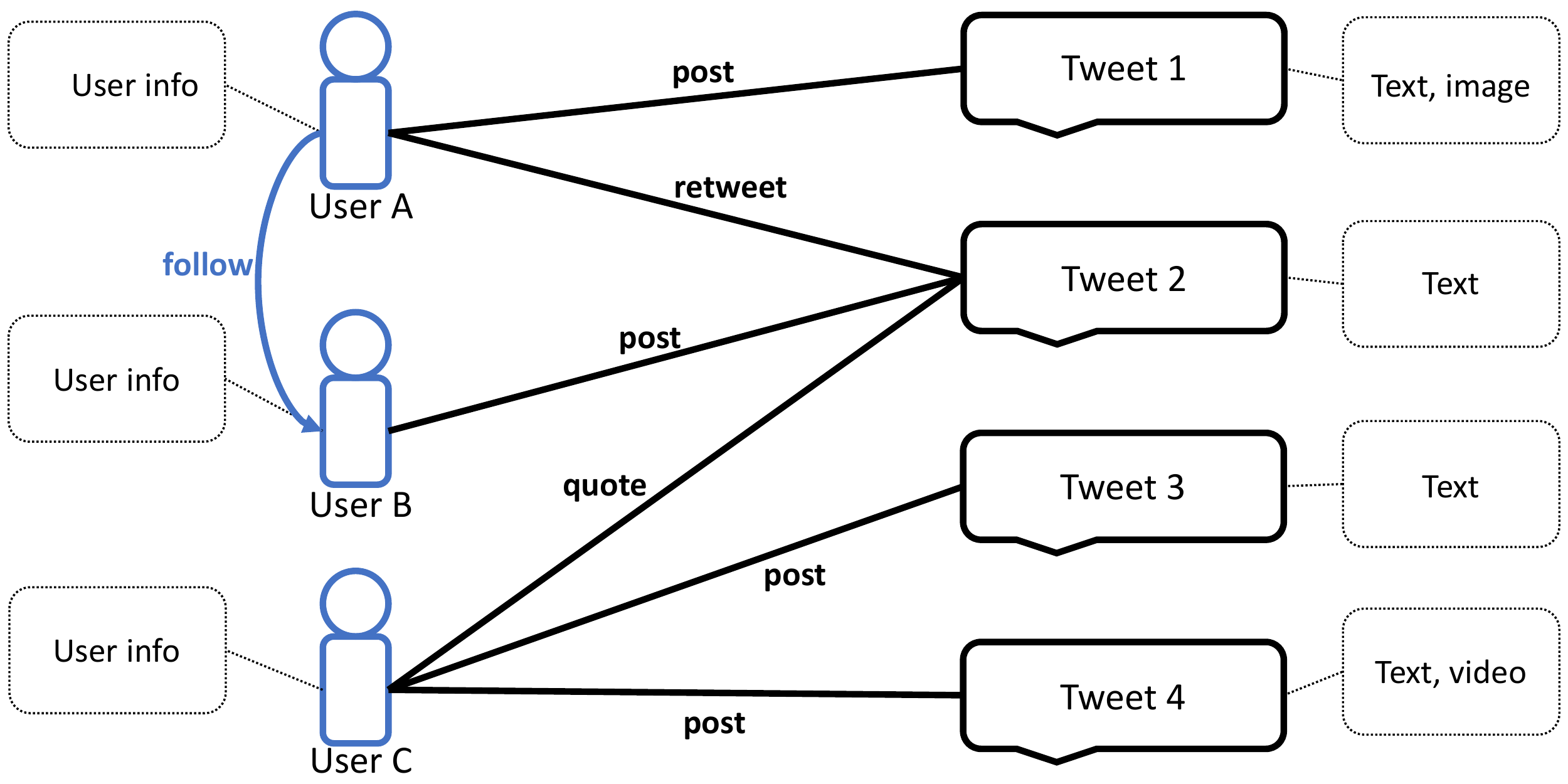}
    \caption{An example of different node types and relation types on Twitter. It is particularly noteworthy that our framework does not need user-user following relations. Collecting the user-user following relations requires an extra time-consuming process. Consequently, without the following relations, the general heterogeneous network is effectively reduced to a {\it bipartite} heterogeneous graph. }
    \label{fig:network_example}
\end{figure}

An example of different node types and relation types on Twitter is illustrated in Figure~\ref{fig:network_example}. The network is composed of two types of nodes - users and tweets. The relations between a user and a tweet include post, retweet, quote, \textit{etc}. A user can follow another user and they can also follow each other. The tweet node has textual information and sometimes it has other media information such as video, image, and link. The user node has user characteristics such as number of {\tt followers}, verified status, and profile description. The heterogeneity of information and complex relations among the nodes make it challenging to uncover insight. Recently, representing graph nodes in a low dimensional vector space based on factorization methods, random walks, and deep learning has been proposed to address these issues~\cite{goyal2018graph}. Graph neural networks have shown promising results in multiple applications including medical image computing~\cite{li2020structured}, rumor detection~\cite{huang2020heterogeneous, lin2021rumor}, sentiment analysis~\cite{huang2019syntax, wang2020relational}, and recommender systems~\cite{song2019session}. More specifically, multiple heterogeneous GNN architectures have been proposed to address the challenge of heterogeneity~\cite{10.1145/3447548.3467350, zhang2019heterogeneous}. Motivated by the need to handle heterogeneity, we adopt a heterogeneous GNN architecture to effectively fuse the multimodal features of heterogeneous nodes and explicitly model the user-item relations in social networks. We then apply our framework to ideology detection and polarization understanding tasks on social platforms.

\section{Material and Method}
\subsection{Datasets}
Politically polarized opinions have been observed in online debates with respect to multiple areas such as economy and public health~\cite{lyu2022social}. We employ two large-scale Twitter datasets that record such online discussions to demonstrate the effectiveness of our framework in learning user embeddings without any human annotated ideology labels. For simplicity, we refer to these two datasets as the \textit{inflation} dataset and \textit{vaccine} dataset in the remainder of this paper. Table~\ref{tab:dataset_stats} shows the descriptive statistics of the datasets. The total size (N = 11,038) of these two datasets is significantly larger than most public datasets for ideology detection (e.g., N = 213 for \citet{chen2017opinion}, N = 2,976 for \citet{xiao2020timme}). For more details on data collection, please refer to Appendix where we also describe how we obtain the pseudo ``ground truth'' political ideology of these social media users. Note that these ``ground truth'' labels are only used for evaluation and \textbf{\textit{not}} used during representation learning.

The objects collected are tweets. Each tweet is associated with multiple fields such as tweet ID, tweet content, and the information of the user who interacts with it (e.g., post, retweet, quote), including user ID, and user's social capital (e.g., number of {\tt followers}). Moreover, if a user $A$ retweets/quotes the tweet of another user $B$, then the tweet information of the original tweet and the user characteristics of the original user (i.e., $B$) will also be associated with this retweet/quote. Hence, the retweet/quote relations are implied during the data collection process. In contrast, user-user following relations are not obtained during this phase. There are two challenges in collecting them. First, an extra data collection process is required, which is prohibitively time-consuming. On Twitter, the number of users a user is following is referred to as the number of {\tt friends}. The average number of {\tt friends} of the users in the \textit{inflation} and \textit{vaccine} datasets are 4,688 (SD = 12,411) and 7,822 (SD = 17,789). Using the Twitter API, it takes \textbf{9 days and 14 hours} to collect all the user-user following relations for the \textit{inflation} dataset, and \textbf{3 days and 21 hours} for the \textit{vaccine} dataset. As of February 2022, there are millions of daily active users on Twitter.\footnote{\url{https://www.omnicoreagency.com/twitter-statistics/} [Accessed 2022-04-05]} It is difficult to collect all the user-user following relations of the users-of-interest in a short period. Second, following relations are sometimes unavailable due to the privacy settings of Twitter users. Therefore, to increase the applicability of our framework and to conduct downstreaming tasks such as ideology detection and polarization analysis in a more timely manner, we \textbf{\textit{do not}} crawl user-user following relations.
%We need to collect the user information of all the users that a user is following to obtain the complete user-user following relations of this user. 

\begin{table}[htbp]
\small
    \centering
    \caption{Statistics of the \textit{inflation} and \textit{vaccine} datasets.}
   
    \begin{tabular}{lccc}
    \toprule
     Dataset & \# unique users & \# unique tweets & \# tweets  \\
     \midrule
    \textit{inflation} & 8,824 &  22,661 & 42,297\\
    % \hline
    \textit{vaccine} & 2,214 &  10,998 & 20,331\\
    \bottomrule
    \end{tabular}
    
    \label{tab:dataset_stats}
\end{table}

\subsection{Graph Representation Learning}
We model the user characteristics, multimodal contents, and the user-item relations in a bipartite heterogeneous graph. We discuss our framework in the following sections.

\subsubsection{Graph definition.}
In the graph $G = (V, E, X)$ of our study, $G$ denotes the graph, $V$ denotes the set of nodes, $E$ denotes the set of edges, and $X$ represents the attributes of a node. There are two types of nodes including users $V_{user} \subset V$ and tweets $V_{tweet} \subset V \:(V_{user} \cup V_{tweet} = V, V_{user} \cap V_{tweet} = \varnothing)$ as shown in Figure~\ref{fig:network_example}. An edge $e \in E$ between a user node $v_{user} \in V_{user}$ and a tweet node $v_{tweet} \in V_{tweet}$ is created if this user interacts with this tweet. In our study, the interactions between a user and a tweet include posting, retweeting and quoting. Since we do not consider the user-user following relations, there is no edge between two user nodes. The retweet and quote relations are represented as a star topology where the tweet that is retweeted/quoted by multiple users serves as the hub of the topology and is connected to these users. No edge is drawn between tweet nodes as well. As a result, the general heterogeneous network is effectively reduced to a \textit{bipartite} heterogeneous graph. The attributes $X_{user} \subset X$ of the user nodes are user characteristics including the number of {\tt followers}, {\tt friends}, {\tt listed memberships}, {\tt statuses}, {\tt favorites}, verification status, as well as the profile description. These publicly available user characteristics are crawled during the data collection. The attributes $X_{tweet} \subset X$ of the tweet nodes have two components: (1) multimodal contents $M_{tweet}$ and (2) the user characteristics of the original author $X_{author}$ \:($X_{tweet} = \{M_{tweet}, X_{user_{author}}\}$). Multimodal contents such as text including hashtags, and image (if any) are used. The user characteristics of the original author of each tweet node are also used to embed the tweet node. In this way, the user-tweet relations are modeled explicitly as each tweet is associated with a vector indicating the information of the original author. 

\subsubsection{Feature embeddings.}
Textual and visual feature embedding methods are employed to capture the semantic meanings of the user and tweet nodes. We use the state-of-the-art framework - Sentence-BERT~\cite{reimers-2019-sentence-bert}, to represent the user profile description and tweet text content. Similar to word embeddings~\cite{mikolov2013efficient}, it converts sentences into a vector space where the vectors of two semantically similar sentences will be closer. The output vectors have 384 dimensions. Inspired by \citet{zhang2018become}, an image is encoded into a feature vector space using ResNet-50~\cite{he2016deep}, which is pre-trained on the ImageNet dataset~\cite{krizhevsky2012imagenet}. The 2048-dimensional image features from the last pooling layer are extracted. In addition, if a tweet is associated with videos, they can be mapped into the feature space via different feature embedding methods. The framework is the same. In our study, we focus on text and image.
% \vspace{-0.1cm}
\subsubsection{Representation learning.}
\begin{figure}[t]
    \centering
    \includegraphics[width = \linewidth]{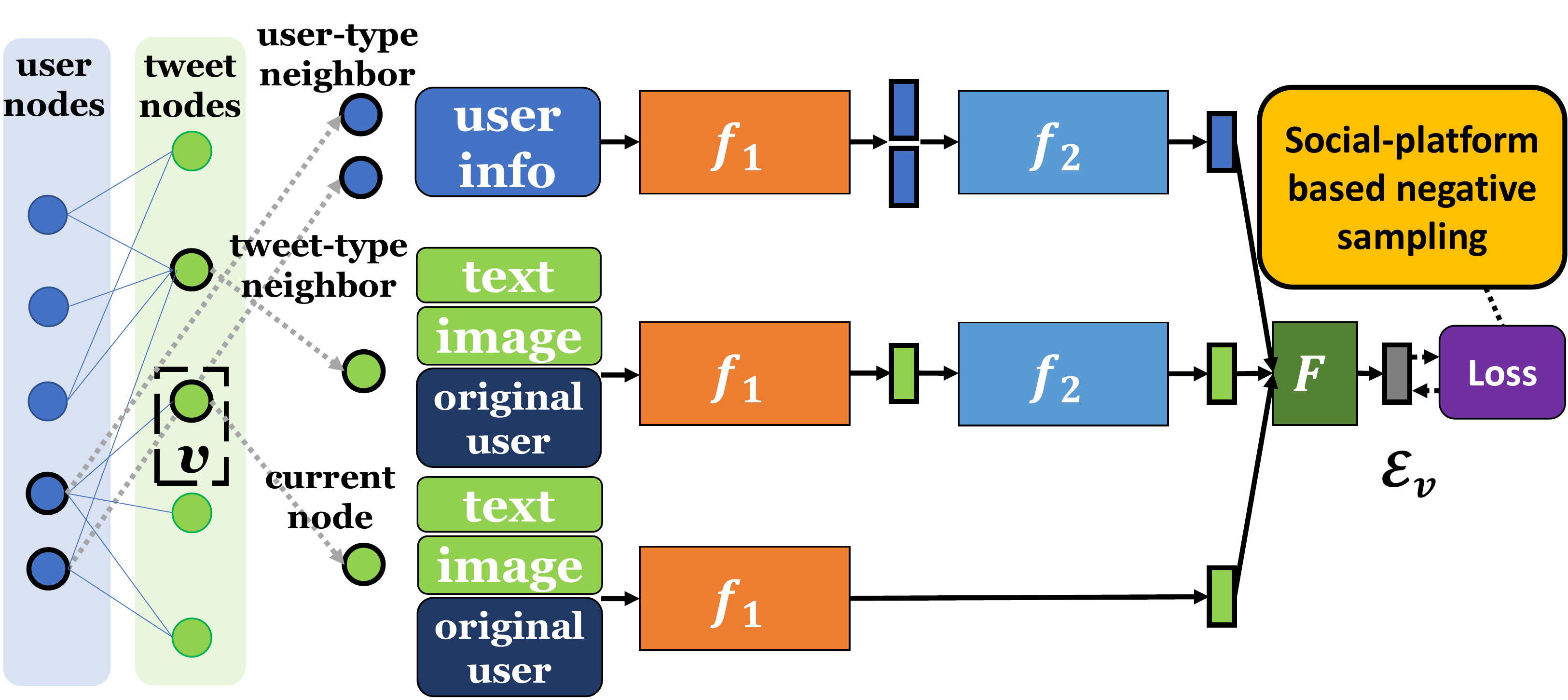}
    % \vspace{-0.6cm}
    \caption{The diagram of our framework.}
    \label{fig:diagram}
\end{figure}

To capture the heterogeneous attributes or contents associated with each node, we adopt the Heterogeneous Graph Neural Network (HetGNN)~\cite{zhang2019heterogeneous} which first uses random walks with restart to generate neighbors for nodes and capture the structural information, and then leverages bi-directional LSTM (Bi-LSTM)~\cite{hochreiter1997long} and the attention mechanism~\cite{velivckovic2018graph} to aggregate node features within each type and among types. Figure~\ref{fig:diagram} illustrates our framework.

% \vspace{-0.4cm}
% \vspace{-0.4cm}

We first fuse heterogeneous attributes $X_{v}$ of node $v \in V$ via a neural network $f_{1}$. In our study, the attributes $X_{user_{u}}$ of the user node $v_{user_{u}} \in V_{user}$ include the number of {\tt followers} and profile description, \textit{etc}. The attributes $X_{tweet_{t}}$ of the tweet node $v_{tweet_{t}} \in V_{tweet}$ include both the multimodal contents $M_{tweet_{t}}$ and the user characteristics $X_{user_{author}}$ of the author who originally posts this tweet ($X_{tweet_{t}} = M_{tweet_{t}} \cup X_{user_{author}}$). Let $\boldsymbol{x_{i}} \in \mathbb{R}^{d_{f} \times 1}$ denote the \textit{i}-th attribute in $X_{v}$ where $d_{f}$ is the feature dimension. The multimodal information can be obtained using different feature embedding methods as aforementioned. Therefore, the attribute feature embeddings $f_{1}(v) \in \mathbb{R}^{d \times 1}$ ($d$: attribute embedding dimension) of $v$ are calculated as follows:
\begin{equation}
    f_{1}(v)=\frac{\sum_{i \in X_{v}} [\overrightarrow{LSTM}\{h(\boldsymbol{x_{i}})\} \oplus \overleftarrow{LSTM}\{h(\boldsymbol{x_{i}})\}]}{|X_{v}|}
\end{equation}
It is noteworthy that $h$ denotes a feature transformer which can be identity or fully connected neural network, \textit{etc}. 

Next, we aggregate the neighbor information of node $v$ via two steps: (1) using Bi-LSTM to aggregate embeddings of the neighbor nodes of the same type, and (2) aggregating the embeddings of the neighbor nodes of different types through the attention mechanism. By employing the random walk with restart strategy, for each node $v \in V$, we generate the user-type sampled neighbor set $N_{user}(v)$ and the tweet-type sampled neighbor set $N_{tweet}(v)$. The aggregated t-type neighbor embeddings $f_{2}^{t}(v) \in \mathbb{R}^{d \times 1}, t \in \{user, tweet\}$ ($d$: aggregated attribute embedding dimension) are then calculated as follows:
\begin{equation}
    f_{2}^{t}(v)=\frac{\sum_{v' \in N_{t}(v)} [\overrightarrow{LSTM}\{f_{1}(v')\} \oplus \overleftarrow{LSTM}\{f_{1}(v')\}]}{|N_{t}(v)|}
\end{equation}
where $v'$ denotes the neighbor nodes in the t-type sampled neighbor set $N_{t}(v)$. The output embeddings $\mathcal{E}_{v} \in \mathbb{R}^{d \times 1}$ of node $v$ are computed as follows:
\begin{equation}\label{eqn:output}
    \mathcal{E}_{v}=\alpha_{v,v}f_{1}(v)+\alpha_{v, user} f_{2}^{user}(v)+\alpha_{v, tweet} f_{2}^{tweet}(v)
\end{equation}
where $\alpha_{v,\cdot}$ represents the attention coefficients of (1) content embeddings of $v$ or (2) the aggregated neighbor embeddings to node $v$. These two sets of embeddings are grouped and denoted as $F(v,i) \in \mathbb{R}^{d \times 1}$ which is defined as follows:
\begin{equation}
    F(v,i)=
    \begin{cases}
    f_{1}(v) & i=v\\
    f_{2}^{t}(v) & i=t \text{, where } t \in\{user, tweet\}
    \end{cases}
\end{equation}
\begin{equation}
    i \in \boldsymbol{P} \text{, where } \boldsymbol{P}=\{v\} \cup \{user, tweet\}
\end{equation}

Therefore, Eq.~(\ref{eqn:output}) is re-formulated as:

\begin{equation}
    \mathcal{E}_{v}=\sum_{i \in \boldsymbol{P}} \alpha_{v,i} F(v,i)
\end{equation}
The attention coefficients $\alpha_{v,i}$ are built by a single-layer feedforward neural network parametrized by a weight vector $\boldsymbol{u} \in \mathbb{R}^{2d \times 1}$, with the LeakyReLU nonlinearity. To make the coefficients comparable, we normalize them across all choices of $i$ using the softmax function. Thus, the coefficients are expressed as:
\begin{equation}
    \alpha_{v,i}=\frac{\exp(\text{LeakyReLU}(\boldsymbol{u^{T}}[F(v,v) \oplus F(v,i)])  )}{\sum_{j \in \boldsymbol{P}}\exp(\text{LeakyReLU}(\boldsymbol{u^{T}}[F(v,v) \oplus F(v,j)])   )}
\end{equation}
where $\boldsymbol{\cdot^{T}}$ represents transposition and $\oplus$ is concatenation. %operation.

We do not require any ideology-related ``supervision'' (i.e., ideology labels) during the training process. The goal of the representation learning of our framework is to maximize the similarity between the embeddings of two nodes if they have similar attribute features and/or are geometrically closer to each other in the graph, otherwise, minimize the similarity. Therefore, we apply the negative sampling technique~\cite{mikolov2013efficient}. The goal can be interpreted as, for each $v \in V$, to minimize the cross entropy loss as follows:
\begin{equation}
    log\sigma(\mathcal{E}_{v_{pos}} \cdot \mathcal{E}_{v}) + log\sigma(-\mathcal{E}_{v_{neg}} \cdot \mathcal{E}_{v})
\end{equation}
where $\sigma$ represents the Sigmoid function, $v_{pos}$ and $v_{neg}$ are the positive and negative sample nodes. \citet{zhang2019heterogeneous} defines the positive sample nodes of the node $v$ as the nodes that can be reached by node $v$ via a random walk, and the negative sample nodes of the node $v$ as any nodes in the graph. Inspired by the findings of \citet{ying2018graph} that importance-based neighborhoods can help improve the representation learning of GNNs, we propose a social-platform based negative sampling strategy. We call it social-platform based because it is based on our observation of the online social platforms, in this case, Twitter. We attempts to measure the likelihood of a user interacting with a tweet. If the likelihood is high, but the user never interacts with the tweet, then this is considered as an important negative pair. The social-platform based negative sampling strategy is designed to capture this kind of negative pairs.

More specifically, we sample the nodes based on the social capital features. This component is only considered when the node type of the negative sample is $user$. If the node type of the negative sample is $tweet$, we apply random sampling. The higher the normalized number of {\tt statuses} a user has posted since the creation of the Twitter account, the more chance the user will be sampled. The motivation is that the number of {\tt statuses} indicates the probability of a user's posting/retweeting/quoting behavior. In other words, it indicates the probability of a user interacting with a tweet. 

We acknowledge that this strategy has limitations. For example, we aggregate the number of {\tt statuses} of each user since the creation of the Twitter account. This implicitly assumes that the user behavior does not change significantly over time. Although it would be better to measure the user behavior in a more fine-grained and more dynamic fashion, it is beyond the scope of this study and left to future work.

Similarly to \citet{zhang2019heterogeneous}, the dimension of the final feature space is set to be 128. For random walk, the walk length is 30 and window size is 5. 

While there are other feature embedding methods and heterogeneous graph representation learning methods that could potentially improve the expressive power of the overall representation~\cite{10.1145/3447548.3467350}, our ultimate goal is not to perfectly characterize the users, tweet contents and relations. Instead, we aim to show that jointly modeling the information in a bipartite heterogeneous graph can benefit the ideology detection task and help understand online political polarization.

% \vspace{-1.2cm}

\begin{table*}[t]
\centering
\small
\caption{Results using the logistic regression model (The best results are highlighted in bold).}
% \vspace{-0.3cm}

\begin{tabular}{lllllll}
\toprule
Dataset                    & \textbf{Representation}      & Accuracy               & Precision              & Recall                 & F1                     & AUROC                  \\ \midrule
\multirow{9}{*}{\textit{Inflation}} & User Info                    & 0.65 +/- 0.01          & 0.38 +/- 0.08          & 0.03 +/- 0.01          & 0.06 +/- 0.01          & 0.50 +/- 0.00          \\  
                           & Textual                      & 0.71 +/- 0.01          & 0.61 +/- 0.03          & \textbf{0.68 +/- 0.03}          & 0.64 +/- 0.02          & 0.70 +/- 0.01          \\ 
                           & Visual & 0.55 +/- 0.09 & 0.44 +/- 0.06 & 0.33 +/- 0.41 & 0.24 +/- 0.21 & 0.51 +/- 0.00 \\

                           & Textual + Visual             &    0.70 +/- 0.01                    &       0.60 +/- 0.03                 &        0.68 +/- 0.04                &       0.64 +/- 0.02                 &           0.70 +/- 0.01             \\  
                           & User Info + Textual + Visual &    0.64 +/- 0.02                    &     0.53 +/- 0.03                   &      0.60 +/- 0.03                  &          0.56 +/- 0.02              &     0.63 +/- 0.02                     \\ 
                           & Late fusion & 0.66 +/- 0.04 & 0.60 +/- 0.07 & 0.51 +/- 0.19 & 0.52 +/- 0.08 & 0.63 +/- 0.04 \\
                           & GCN~\cite{kipf2016semi} & 0.72 +/- 0.02 & 0.71 +/- 0.03 & 0.32 +/-  0.02 & 0.44 +/- 0.02 & 0.62 +/- 0.01\\
                           
                           & GAT~\cite{velivckovic2018graph} & 0.74 +/-  0.02 & 0.70 +/- 0.03 & 0.45 +/- 0.02 & 0.55 +/- 0.02 & 0.67 +/- 0.01\\

                           & MBPHGNN                      & \textbf{0.83 +/- 0.01} & \textbf{0.84 +/- 0.02} & 0.64 +/- 0.02 & \textbf{0.73 +/- 0.02} & \textbf{0.79 +/- 0.01} \\ \hhline{=======}
\multirow{9}{*}{\textit{Vaccine}}   & User Info                    & 0.84 +/- 0.02          & 0.78 +/- 0.05          & 0.58 +/- 0.04          & 0.67 +/- 0.04          & 0.76 +/- 0.02          \\  
                           & Textual                      & 0.78 +/- 0.03          & 0.74 +/- 0.07          & 0.28 +/- 0.07          & 0.40 +/- 0.07          & 0.62 +/- 0.03          \\ 
                           & Visual & 0.73 +/- 0.03 & 0.35 +/- 0.45 & 0.01 +/- 0.01 &0.02 +/- 0.02 &0.50 +/- 0.01\\

                           & Textual + Visual             &    0.78  +/- 0.02          &        0.76  +/- 0.10                &      0.28  +/- 0.05              &   0.40  +/- 0.05                 &     0.62  +/- 0.02             \\
                           & User Info + Textual + Visual &     0.82  +/- 0.04                   &         0.66  +/- 0.09               &    0.66  +/- 0.05                    &         0.66  +/- 0.07               &     0.77  +/- 0.04                   \\ 
                           & Late fusion & 0.77 +/- 0.03 & 0.87 +/- 0.20 & 0.30 +/- 0.16 & 0.39 +/- 0.10 & 0.63 +/- 0.05\\ 
                           & GCN~\cite{kipf2016semi} & 0.72 +/- 0.02 & 0.39 +/- 0.39 & 0.02 +/- 0.02 & 0.03 +/- 0.03 & 0.50 +/- 0.01\\
                           & GAT~\cite{velivckovic2018graph} & 0.72 +/- 0.02 & 0.23 +/- 0.20 & 0.01 +/- 0.01 & 0.03 +/- 0.02 & 0.50 +/- 0.01\\
                           
                           & MBPHGNN                      & \textbf{0.92 +/- 0.02} & \textbf{0.91 +/- 0.05} & \textbf{0.80 +/- 0.05} & \textbf{0.85 +/- 0.05} & \textbf{0.88 +/- 0.03} \\ \bottomrule
\end{tabular}

\label{tab:experiments_lr}
\end{table*}

\begin{table*}[t]
\centering
\small
\caption{Results using the random forest model (The best results are highlighted in bold).}
% \vspace{-0.3cm}

\begin{tabular}{lllllll}
\toprule
Dataset                    & \textbf{Representation}               & Accuracy               & Precision              & Recall                 & F1                     & AUROC                  \\ \midrule
\multirow{7}{*}{\textit{Inflation}} & User Info                    & 0.65 +/- 0.01          & 0.34 +/- 0.12          & 0.02 +/- 0.01          & 0.04 +/- 0.01          & 0.50 +/- 0.00          \\
                           & Textual                      & 0.70 +/- 0.02          & 0.60 +/- 0.04          & 0.67 +/- 0.04          & 0.63 +/- 0.03          & 0.69 +/- 0.02          \\ 
                           & Visual & 0.55 +/- 0.09 & 0.45 +/- 0.09 & 0.34 +/- 0.41 & 0.25 +/- 0.20 & 0.51 +/- 0.01\\  
                           
                           & Textual + Visual             &    0.69 +/- 0.02                     & 0.59 +/- 0.04                       &         \textbf{0.68 +/- 0.03}              &         0.63 +/- 0.03               &       0.69 +/- 0.02     \\ 
                           & User Info + Textual + Visual &    0.64 +/- 0.02                    &     0.53 +/- 0.03                   &      0.60 +/- 0.03                  &          0.56 +/- 0.02              &     0.63 +/- 0.02                   \\ 
                           & Late fusion & 0.62 +/- 0.02 & 0.54 +/- 0.07 & 0.11 +/- 0.02 & 0.18 +/- 0.04 & 0.53 +/- 0.01 \\  
                           & GCN~\cite{kipf2016semi} & 0.70 +/- 0.02 & 0.64 +/- 0.01 & 0.32 +/- 0.03 & 0.42 +/- 0.02 & 0.61 +/- 0.01\\
                           & GAT~\cite{velivckovic2018graph} & 0.71 +/- 0.01 & 0.67 +/- 0.04 & 0.32 +/- 0.02 & 0.43 +/- 0.02 & 0.62 +/- 0.01\\
                           & MBPHGNN                      & \textbf{0.84 +/- 0.01} & \textbf{0.87 +/- 0.02} & 0.63 +/- 0.02 & \textbf{0.73 +/- 0.02} & \textbf{0.79 +/- 0.01} \\ 
                           \hhline{=======}
\multirow{9}{*}{\textit{Vaccine}}   & User Info                    & 0.81 +/- 0.02          & 0.88 +/- 0.06          & 0.34 +/- 0.05          & 0.48 +/- 0.06          & 0.66 +/- 0.03          \\  
                           & Textual                      & 0.74 +/- 0.03          & 0.80 +/- 0.29          & 0.07 +/- 0.05         & 0.12 +/- 0.08          & 0.53 +/- 0.02          \\ 
                           & Visual & 0.73 +/- 0.03 & 0.40 +/- 0.49 & 0.01 +/- 0.01 & 0.02 +/- 0.03 & 0.51 +/- 0.01  \\

                           & Textual + Visual             &      0.74  +/- 0.03                  &       0.68 +/- 0.25                 &    0.08 +/- 0.04                   &        0.14  +/- 0.07                &      0.53  +/- 0.02                  \\  
                           & User Info + Textual + Visual &    0.79  +/- 0.04                    &        0.84  +/- 0.11                &    0.25 +/- 0.07                    &        0.38  +/- 0.09                &     0.62  +/- 0.04                   \\  
                           & Late fusion & 0.75 +/- 0.04 & 0.86 +/- 0.30 & 0.09 +/- 0.11 & 0.14 +/- 0.15 & 0.54 +/- 0.05 \\ 
                           & GCN~\cite{kipf2016semi} & 0.73 +/- 0.02 & 0.47 +/- 0.26 & 0.03 +/- 0.02 & 0.06 +/- 0.03 & 0.51 +/- 0.01\\
                           & GAT~\cite{velivckovic2018graph} & 0.72 +/- 0.02 & 0.44 +/- 0.08 & 0.14 +/- 0.02 & 0.21 +/- 0.03 & 0.53 +/- 0.01\\
                           
                           & MBPHGNN                      & \textbf{0.93 +/- 0.02} & \textbf{0.94 +/- 0.05} & \textbf{0.77 +/- 0.05}          & \textbf{0.85 +/- 0.04} & \textbf{0.88 +/- 0.03} \\ \bottomrule
\end{tabular}

\label{tab:experiments_rf}
\end{table*}

% (left/right: 66\%/34\%)  (left/right: 73\%/27\%)
\section{Results}
\subsection{Political Ideology Detection}
We refer to our framework as MBPHGNN (Multimodal Bipartite Heterogeneous Graph Neural Network). To evaluate the quality of the embeddings learned via MBPHGNN, we compare the performance on a user ideology detection task on two aforementioned Twitter datasets using the learned embeddings and multiple other embeddings that are associated with the users. The ``ground truth'' ideology label is obtained as discussed in Appendix. If the political score is greater than or equal to zero, the label is right-leaning (accounts for 34\%/27\% in \textit{inflation}/\textit{vaccine}), otherwise it is left-leaning (accounts for 66\%/73\% in \textit{inflation}/\textit{vaccine}). Ideology detection is a binary classification task. We expect our models to predict whether a user is left- or right-leaning. The embeddings are fed into a logistic regression classifier and a random forest classifier. For the logistic regression model, the elastic net regularization is used ($l1\_ratio=1$, $C=0.5$). For the random forest model, the number of trees is set to 100. The performance is measured using 10-fold cross-validation. 

We compare MBPHGNN with the following state-of-the-art embedding methods to demonstrate the effectiveness of our framework over unimodal frameworks, early/late fusion strategies, and homogeneous GNN frameworks. The GCN~\cite{kipf2016semi} and GAT~\cite{velivckovic2018graph} baselines consider both user and tweet nodes the same type. The user and multimodal information is preprocessed using the well-studied principle component analysis~\cite{wold1987principal}. The InfoNCE loss~\cite{oord2018representation} is used to train the GCN and GAT baselines. The embeddings we compare include:

\begin{itemize}
    \item \textbf{User info:} The concatenation of the number of {\tt followers}, {\tt friends}, {\tt listed memberships}, {\tt statuses}, {\tt favorites}, verification status and the sentence embeddings~\cite{reimers-2019-sentence-bert} of the profile description is used.
    \item \textbf{Textual:} Only the sentence embeddings~\cite{reimers-2019-sentence-bert} of the tweet textual content are used. 
    \item \textbf{Visual:} Only the image embeddings~\cite{he2016deep} are used. If a tweet is not associated with an image, the visual embeddings will be filled with zeros. 
    \item \textbf{Textual + Visual:} The concatenation of text and image embeddings is used.  
    \item \textbf{User info + Textual + Visual:} The concatenation of user characteristics, text and image embeddings is used. 
    \item \textbf{Homogeneous GNN:} The embeddings learned by GCN~\cite{kipf2016semi} and GAT~\cite{velivckovic2018graph} are used.
    \item \textbf{TIMME}~\cite{xiao2020timme}\textbf{:} The embeddings learned via the link prediction task of the TIMME framework are used.
\end{itemize}

In addition, we construct a \textbf{Late fusion} model. The predicted ideology label of each user is calculated from the weighted outcomes of \textbf{User Info}, \textbf{Textual} and \textbf{Visual} models. Weights are assigned based on the F1 score of each single model. 

Each user may post multiple tweets. The embeddings of these tweets are fed to the classifier, respectively. The predicted ideology label of this user is voted by the predicted outcome of each tweet. %This strategy is applied if a user is associated with multiple tweets.

Since during representation learning, no ideology labels are provided for the frameworks, the embeddings learned by TIMME~\cite{xiao2020timme} are obtained via the link prediction task. In addition, TIMME does not compute the embeddings of isolated nodes. Therefore, the comparison between MBPHGNN and TIMME is conducted for the set of nodes with at least one neighbour. The experiment results are reported in separate tables.

Tables \ref{tab:experiments_lr} and \ref{tab:experiments_rf} show the ideology detection performance of the logistic regression and the random forest classifiers. We can observe that, our framework performs well, which achieves an overall accuracy of 84\% and 93\%, and outperforms other baselines by a margin of at least \textbf{9\% and 11\% absolute gain} in AUROC on the \textit{inflation} and \textit{vaccine} datasets, respectively. \textbf{Textual+Visual} achieves a higher recall in the \textit{inflation} dataset. After further investigation, we find that it tends to assign the default class label to all users, resulting in a higher recall but a lower precision. It does not perform well in distinguishing users of different groups. F1 and AUC still suggest that the embeddings learned by MBPHGNN are overall better in ideology detection. Further, the aid of visual information by simple concatenation seems inconsistent across two datasets, which might be because concatenation cannot capture the high-level correlations among different modalities. 

Although Twitter allows users to describe themselves in the profile description section, simply reading the user profile only is not reliable enough to detect political ideology as shown in the rows of \textbf{User Info} in Tables \ref{tab:experiments_lr} and \ref{tab:experiments_rf}. The overall accuracy is around 65\% and 84\% in the \textit{inflation} and \textit{vaccine} datasets. The low performance of the GCN and GAT frameworks, which consider the user and tweet nodes as the same type, suggests the importance of addressing heterogeneity. Interestingly, the choice of classifier does not make much difference on the \textit{inflation} dataset. On the other hand, except for MBPHGNN, the performance of random forest is worse than logistic regression on the \textit{vaccine} dataset, which might be related to the number of noise variables~\cite{kirasich2018random}, again suggesting that the embeddings leanred via MBPHGNN are more robust. 

Tables~\ref{tab:experiments_lr_timme} and \ref{tab:experiments_rf_timme} show that MBPHGNN achieves a better performance than TIMME~\cite{xiao2020timme}. More importantly, we find that the prediction performance of MBPHGNN on connected nodes is better than the prediction performance on isolated nodes (e.g., Table~\ref{tab:experiments_lr} versus Table~\ref{tab:experiments_lr_timme}), indicating the effect of modeling connections. 

\begin{table*}[htbp]
\centering
\small
\caption{Results using the logistic regression model between TIMME and MBPHGNN (The best results are highlighted in bold).}
% \vspace{-0.3cm}

\begin{tabular}{lllllll}
\toprule
Dataset                    & \textbf{Representation}      & Accuracy               & Precision              & Recall                 & F1                     & AUROC                  \\ \midrule
\multirow{2}{*}{\textit{Inflation}} 
                          & TIMME~\cite{xiao2020timme} & 0.90 +/- 0.02 & 0.86 +/- 0.02 & 0.85 +/- 0.03& 0.86 +/- 0.02 & 0.89 +/- 0.02\\

                           & MBPHGNN                      & \textbf{0.90 +/- 0.02} & \textbf{0.87 +/- 0.02} & \textbf{0.87 +/- 0.04} & \textbf{0.87 +/- 0.02} & \textbf{0.90 +/- 0.02} \\ \hhline{=======}
\multirow{2}{*}{\textit{Vaccine}}   
                           & TIMME~\cite{xiao2020timme} & 0.96 +/- 0.02 & 0.93 +/- 0.04 & 0.92 +/- 0.04 & 0.92 +/- 0.03 & 0.95 +/- 0.02\\
                           
                           & MBPHGNN                      & \textbf{0.96 +/- 0.01} & \textbf{0.93 +/- 0.04} & \textbf{0.93 +/- 0.02} & \textbf{0.93 +/- 0.02} & \textbf{0.95 +/- 0.01} \\ \bottomrule
\end{tabular}

\label{tab:experiments_lr_timme}
\end{table*}

\begin{table*}[htbp]
\small
\centering
\caption{Results using the random forest model between TIMME and MBPHGNN (The best results are highlighted in bold).}
% \vspace{-0.3cm}

\begin{tabular}{lllllll}
\toprule
Dataset                    & \textbf{Representation}               & Accuracy               & Precision              & Recall                 & F1                     & AUROC                  \\ \midrule
\multirow{2}{*}{\textit{Inflation}} 
                           & TIMME~\cite{xiao2020timme} & 0.90 +/- 0.01 & 0.87 +/- 0.02 & 0.84 +/- 0.03 & 0.85 +/- 0.02 & 0.88 +/- 0.02\\
                           & MBPHGNN                      & \textbf{0.91 +/- 0.01} & \textbf{0.88 +/- 0.02} & \textbf{0.87 +/- 0.04} & \textbf{0.87 +/- 0.02} & \textbf{0.90 +/- 0.02} \\ 
                           \hhline{=======}
\multirow{2}{*}{\textit{Vaccine}}   
                           & TIMME~\cite{xiao2020timme} & 0.96 +/- 0.02 & 0.94 +/- 0.04 & 0.91 +/- 0.05 & 0.92 +/- 0.04 & 0.94 +/- 0.03\\
                           
                           & MBPHGNN                      & \textbf{0.96 +/- 0.01} & \textbf{0.95 +/- 0.04} & \textbf{0.91 +/- 0.03}          & \textbf{0.93 +/- 0.03} & \textbf{0.95 +/- 0.02} \\ \bottomrule
\end{tabular}

\label{tab:experiments_rf_timme}
\end{table*}

%Further, the aid of visual information by simple concatenation seems inconsistent across two datasets, which might be because concatenation cannot capture the high-level correlations among different modalities. %However, in the following section, we will show that our framework is able to model the multimodal contents to learn meaningful embeddings. 

% \begin{figure*}[t]
%     \centering
%     \includegraphics[width = \linewidth]{six_com.pdf}
%     % \vspace{-0.5cm}
%     \caption{\textbf{t-SNE visualization of the output of six different embeddings (upper: \textit{inflation}, lower: \textit{vaccine}).} The color indicates the political scores. The bluer the color is, the lower the political score is (more left-leaning). The redder the color is, the higher the political score is (more right-leaning). We highlight certain areas in Panels (f) and (l) for further analysis on political polarization.
%     }
%     \label{fig:six_com}
% \end{figure*}

\begin{figure}[htbp]
    \centering
    \includegraphics[width = \linewidth]{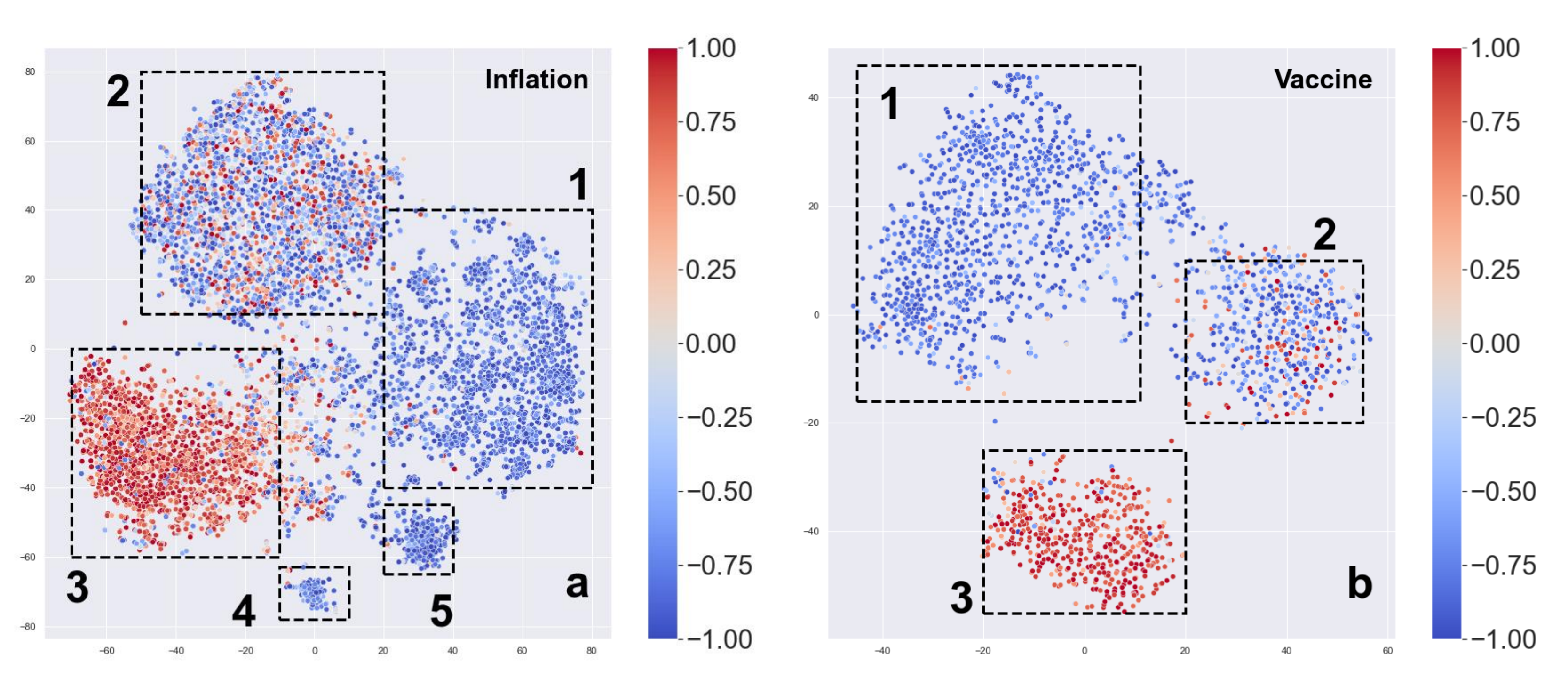}
    % \vspace{-0.5cm}
    \caption{\textbf{t-SNE visualization of the output of MBPHGNN (a: \textit{inflation}, b: \textit{vaccine}).} The color indicates the political scores. The bluer the color is, the lower the political score is (more left-leaning). The redder the color is, the higher the political score is (more right-leaning). We highlight certain areas for further analysis on political polarization.
    }
    \label{fig:six_com}
\end{figure}

To have a more intuitive understanding of the relations between the embeddings and political ideology, we apply t-distributed stochastic neighbor embedding method (t-SNE) which can visualizes high-dimensional data by giving each data point a location in a two or three-dimensional map. t-SNE is capable of capturing the local as well as the global structures to reveal information of the presence of clusters~\cite{van2008visualizing}. Figure~\ref{fig:six_com} shows the t-SNE outputs of the embeddings of MBPHGNN.

%Figure~\ref{fig:six_com} (top: \textit{inflation}, bottom: \textit{vaccine}) shows the t-SNE outputs of the embeddings of the \textbf{User Info}, \textbf{Textual}, \textbf{Visual}, \textbf{Textual + Visual}, \textbf{User Info + Textual + Visual}, and \textbf{MBPHGNN}. The small clusters in Figures~\ref{fig:six_com}a and \ref{fig:six_com}g are the users without descriptions. The small clusters in Figures~\ref{fig:six_com}d and \ref{fig:six_com}j are the tweets with no images.

%As shown in Figure~\ref{fig:six_com}, across two different datasets, the embeddings learned via our framework align the best in terms of the political ideology. There is clear segregation between the red (right-leaning) and blue (left-leaning) clusters. Although we can roughly observe clusters of the same color in \textbf{Textual} and \textbf{Textual + Visual}, they tend to mix with each other, suggesting that only modeling contents and user characteristics is not effective compared with our framework which also considers the user-item relations. 

\begin{figure}[t]
    \centering
    \includegraphics[width = \linewidth]{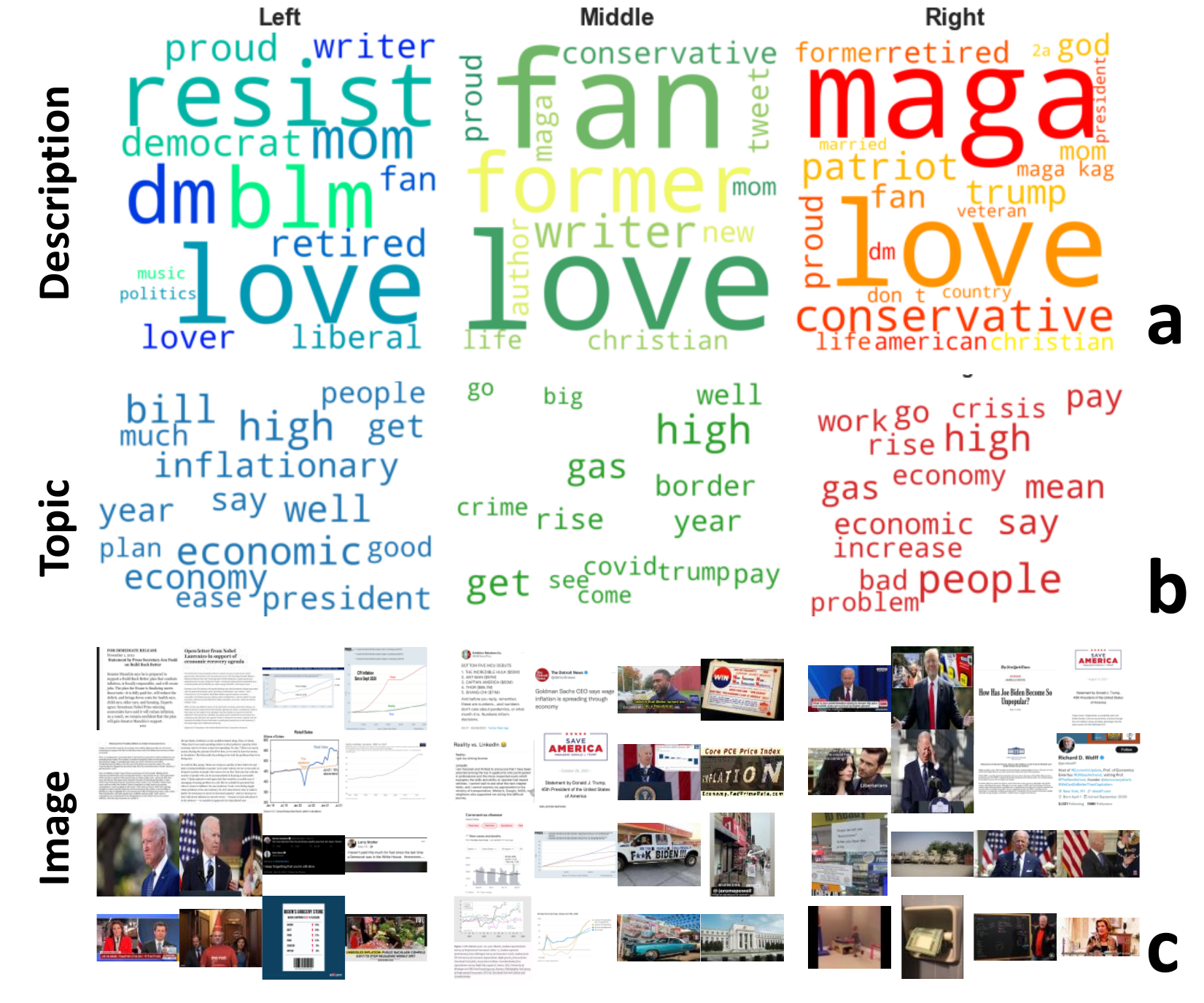}
    % \vspace{-0.3cm}
    \caption{\textbf{User descriptions and the multimodal post contents that are related to the users of {\tt Left}, {\tt Middle}, and {\tt Right} clusters of the \textit{inflation} dataset.} Panel (a) shows the word cloud of the user descriptions of the users in each group. The size of the word is proportional to the frequency of the appearance. Panel (b) shows the word cloud of the most popular topic of each group. The size of the word is proportional to the weight assigned by the LDA model. Panel (c) lists the representative images that are shared by the users of each group. See Appendix for details on the word weights.}
    % \vspace{-0.3cm}
    \label{fig:user_modal_inflation}
\end{figure}
% \vspace{-0.5cm}
\subsection{Political Polarization Understanding}
In this section, we conduct a characterization study to show that MBPHGNN learns meaningful embeddings and can help better understand online political polarization. In Figures~\ref{fig:six_com}a and \ref{fig:six_com}b, we observe that except for the clusters that are predominately blue (Area 1) or red (Area 3), there is a third cluster (Area 2) which is a mixture of the red and blue nodes. For better presentation, we highlight them in Figures~\ref{fig:six_com}a and \ref{fig:six_com}b. For simplicity, we refer to the users of Areas 1, 2, 3 as {\tt Left}, {\tt Middle}, and {\tt Right}, respectively.

\begin{figure}[t]
    \centering
    \includegraphics[width =\linewidth]{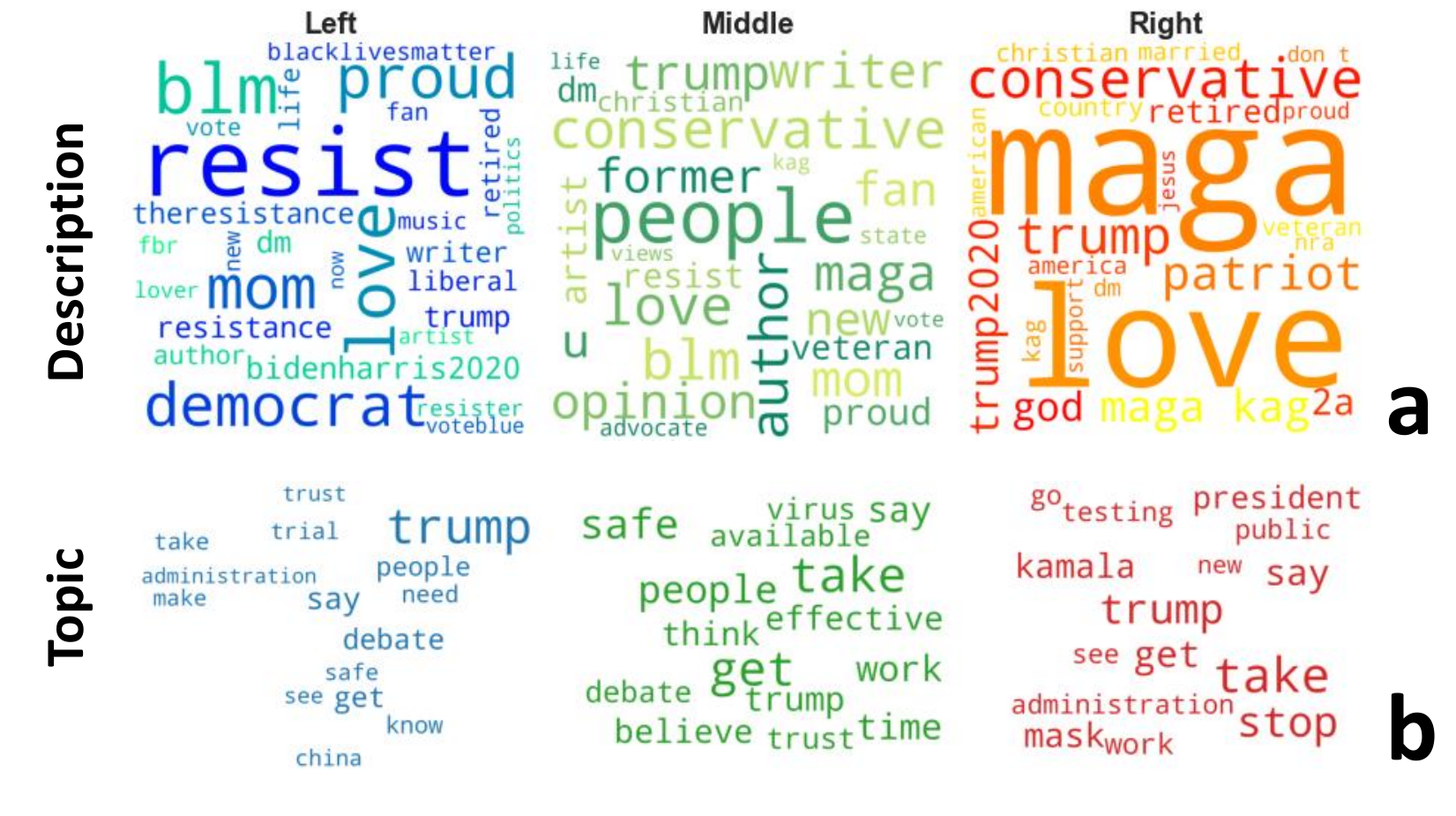}
    % \vspace{-0.3cm}
    \caption{\textbf{User descriptions and the textual contents that are related to the users of {\tt Left}, {\tt Middle}, and {\tt Right} clusters of the \textit{vaccine} dataset.} Panel (a) shows the word cloud of the user descriptions of the users in each group. %The size of the word is proportional to the frequency of the appearance.
    Panel (b) shows the word cloud of the most popular topic of each group. See Appendix for details on the word weights.} 
    \label{fig:user_modal_vaccine}
    % \vspace{-0.2cm}
\end{figure}

\subsubsection{Users with similar politics-related descriptions tend to be in the same cluster.}
To understand the differences among the users of {\tt Left}, {\tt Middle}, and {\tt Right}, we visualize the user descriptions and the multimodal post contents. Figures~\ref{fig:user_modal_inflation}a and ~\ref{fig:user_modal_vaccine}a show the word clouds of the user descriptions of the users in each group of the \textit{inflation} and \textit{vaccine} datasets. Interestingly, ``love'' is observed in the descriptions of all three groups across two datasets. The words of {\tt Left} and {\tt Right} are associated with the political affiliation. For instance, there are ``blm'', ``resist'', and ``democrat'' in {\tt Left}, while ``maga'', ``conservative'', ``trump'', and ``chirstian'' in {\tt Right}. Left-leaning users are found to be more engaged in online civil right movements such as ``\#BlackLivesMatter''~\cite{panda2020covid} and ``\#StopAsianHate''~\cite{lyu2021understanding}. ``resist'' represents the American liberal political movement that protested the presidency of  Trump.\footnote{\url{https://en.wikipedia.org/wiki/The_Resistance_(American_political_movement)} [Accessed 2022-01-09]} ``Make America Great Again'' or ``MAGA'' is a campaign slogan popularized by  Trump.\footnote{\url{https://en.wikipedia.org/wiki/Make_America_Great_Again} [Accessed 2022-01-09]} With respect to ``christian'', \citet{pewresearchcenter_2015_Americansfavor} found there are proportionally more Christians among conservatives than among Democrats and Democratic leaners.
% \vspace{-0.1cm}

\subsubsection{Topics are different across three groups.} 
Figures~\ref{fig:user_modal_inflation}b and \ref{fig:user_modal_vaccine}b show the word clouds of the most popular topics.\footnote{We describe how we determine the most popular topics in Appendix.} We observe that users of different clusters focus on different topics. Take the \textit{inflation} dataset as an example, we find that ``gas'' and ``crisis'' in {\tt Right} are related to the tweets that express negative opinions against President Joe Biden\footnote{The tweet has been paraphrased to protect the user privacy.}:

{\it ``These are what \#BidenDelivers: Soaring gas prices, inflation, millions of unfilled jobs, Crisis in the Middle East, Border issue, Woke Military, Increasing crime, Supply chain issue, Critical Race Theory, Taxes, Refugee resettlement, IRS expansion.''}

\subsubsection{The styles and contents of images imply political ideology.} 
Note that given the limited number of unique images in the \textit{vaccine} dataset, we only conduct analysis on the images of the \textit{inflation} dataset. The images that are related to {\tt Left}, {\tt Middle} and {\tt Right} are in general different. Figure~\ref{fig:user_modal_inflation}c shows the representative figures.\footnote{We discuss how the representative figures are selected in Appendix.} Almost all groups post images of political figures (e.g., Joe Biden, Donald Trump). In {\tt Left} (lower left) and {\tt Middle} (upper right), we observe more portraits, while in {\tt Right} (upper left) there are more sarcasm images of Joe Biden. Interestingly, users of {\tt Right} share more eye-catching images. We find users in {\tt Left} share screenshots of long paragraphs and charts that explain the reasons for inflation (upper left and right in {\tt Left}). Users in {\tt Middle} use images of items that can reflect inflation (lower right in {\tt Middle}) more often. However, users in {\tt Right} use images that intend to depict inflation more vividly (lower left in {\tt Right}).

\subsubsection{The levels of retweet/quote activity are different.} 
We calculate the number of users per unique tweet in each group. This value indicates the number of retweet/quote activities. The higher this value is, the more users retweet/quote the same tweet, suggesting more retweet/quote activities. There are more retweet/quote activities in the {\tt Left} cluster (4.58/1.73 users per unique tweet in \textit{inflation}/\textit{vaccine}) and {\tt Right} cluster (2.90/1.90 users per unique tweet in \textit{inflation}/\textit{vaccine})  than in the {\tt Middle} cluster (1.03/1.02 users per unique tweet in \textit{inflation}/\textit{vaccine}). Fewer retweet/quote activities indicate a sparser network on which GNN normally performs poorly due to the limited knowledge gain from less representative neighbors~\cite{jia2020personalized}.

For the \textit{inflation} dataset, we also observe two sub-clusters composed of left-leaning users (highlighted in Areas 4 and 5 in Figure~\ref{fig:six_com}a). By analyzing the user characteristics and the multimodal contents, we find that it is mainly because of the difference in user descriptions. Unlike previous studies~\cite{waller2021quantifying, conover2011political} that focus on the political polarization in terms of two groups (left and right), by combining information of users, contents, and relations, we are able to conduct a more fine-grained analysis and provide insights into political polarization guided by the clusters found by our ideology-agnostic representation learning framework (i.e., no political labels during training). In our case, the three clusters are different regarding user descriptions, multimodal contents and retweet/quote activities.

\section{Conclusion and Future Work}

We introduce an effective and efficient framework based on bipartite heterogeneous graph neural networks to decode user-content interactions by jointly modeling user characteristics, multimodal contents and user-item relations. The user embeddings learned via our framework help achieve improved performance in an ideology detection task on two social media datasets. In addition, our framework helps obtain a more fine-grained understanding of political polarization on social platforms.

There are a few limitations in our studies. Our study only focuses on the political polarization in the United States. Although the topics of the two datasets we use are diverse, the periods of the data collections are close. We can extend our study in a few directions. First, motivated by the findings of \citet{waller2021quantifying} that social media users became more polarized after 2016, we plan to apply our framework to the online discussions before and after 2016 of more countries (e.g., U.K., Canada, and France) to further investigate political polarization. Second, our framework is designed to decode user-content interactions on user-item bipartite information networks. We plan to apply our framework to other applications, such as content and product recommendation to users.

% \begin{acks}
% We are grateful to Wei Xiong, Weijian Li, Wei Zhu, Dr. Yinglong Xia, Dr. Yan Zhu and Prof. Chen Ding for their constructive suggestions. This research was supported in part by a University of Rochester Research Award.
% \end{acks}
\section{Acknowledgments}
We are grateful to Wei Xiong, Weijian Li, Wei Zhu, Dr. Yinglong Xia, Dr. Yan Zhu and Prof. Chen Ding for their constructive suggestions. This research was supported in part by a University of Rochester Research Award.

\bibliography{sample-base}

\appendix

\section{Supplemental Material and Method}

\subsection{Datasets}
In our study, we employ two large-scale Twitter datasets that record online discussions on economy and public health to demonstrate the effectiveness of our framework in learning user embeddings without any human annotated ideology labels. For the economy topic, we choose to investigate the discussions about inflation during the COVID-19 pandemic as a case study. According to a Yahoo News/YouGov poll,\footnote{\url{https://news.yahoo.com/poll-77-percent-of-americans-now-say-inflation-is-personally-affecting-them-and-57-percent-blame-biden-210739716.html} [Accessed 2022-01-05]} about 90 percent of Republicans assign President Biden at least some blame for inflation while only 28 percent of Democrats do. In terms of public health topic, we focus on the discussions regarding COVID-19 vaccines, since evidence shows that there is an opinion divide in the presence of political affiliations~\cite{lyu2022social}.

This study leverages the publicly available tweets collected using the Twitter API.\footnote{\url{https://www.tweepy.org/} [Accessed 2021-12-05]} We use lists of keywords and hashtags to acquire tweets. To construct the \textit{inflation} dataset, unlike \citet{angelico2021can} who use ``inflation'', ``inflationary'', ``price'', ``expensive'', ``cheap'', and ``expensive bill'' as the search keywords, we only use one keyword - ``inflation'' and one hashtag - ``\#inflation'' to collect the tweets regarding inflation because using their list of keywords may collect many false positive tweets. For example, {\tt  ``Pakistanis will surely pay the price of sending terrorists to Afghanistan.''} would have been collected because of the word ``price''. However, it is apparently not related to the inflation issue. The dataset of \citet{lyu2022social} is used as the \textit{vaccine} dataset in our study. A detailed description of the data collection process can be found in \citet{lyu2022social}. Additionally, our study focuses on the political polarization in the United States. Therefore, we only include the Twitter users that are located in the United States by extracting the location information from public user profiles.

\subsection{Political Ideology}
One of the applications of our framework is to detect the political ideology of social media users. To evaluate the performance, we need the ``ground truth'' labels. However, most ordinary citizens' political ideology is unknown or hidden while human annotated labels would be costly to obtain as aforementioned. Therefore, motivated by the findings of \citet{golbeck2014method} that Twitter users mostly follow accounts that share the same political ideology, we infer each user's political ideology by examining the number of Twitter accounts with known political leaning that a given user follows. Several previous studies on selective exposure to political information suggest that people look for information from people with similar political views~\cite{frey1986recent,garrett2009politically}. The experimental results of both quantitative and qualitative validations also confirm that calculating the political scores based on the following behavior yields an accurate estimate of people's true political leaning~\cite{golbeck2014method}.

The accounts with known political leaning include media accounts, journalists and political figures. The political leanings of the media accounts and the journalists are judged and assigned by allsides.com\footnote{\url{https://www.allsides.com/media-bias/media-bias-ratings} [Accessed 2021-12-05]} and politico.com,\footnote{\url{https://www.politico.com/blogs/media/2015/04/twitters-most-influential-political-journalists-205510} [Accessed 2021-12-05]} respectively. This set of Twitter accounts is roughly balanced with around 400 right-leaning and 400 left-leaning. Prominent politicians are excluded since Twitter users may just follow them because of their popularity and importance. Using the Twitter API, we collect the Twitter accounts that the Twitter users follow and count the number of right- and left-leaning accounts, denoted by $N_{R}$ and $N_{L}$, respectively. The political score $P_{S}$ is calculated as follows:
\begin{equation}
    P_{S}=\frac{N_{R}-N_{L}}{N_{R} + N_{L}}
\end{equation}

\begin{figure}[t]
    \centering
    \includegraphics[width =\linewidth]{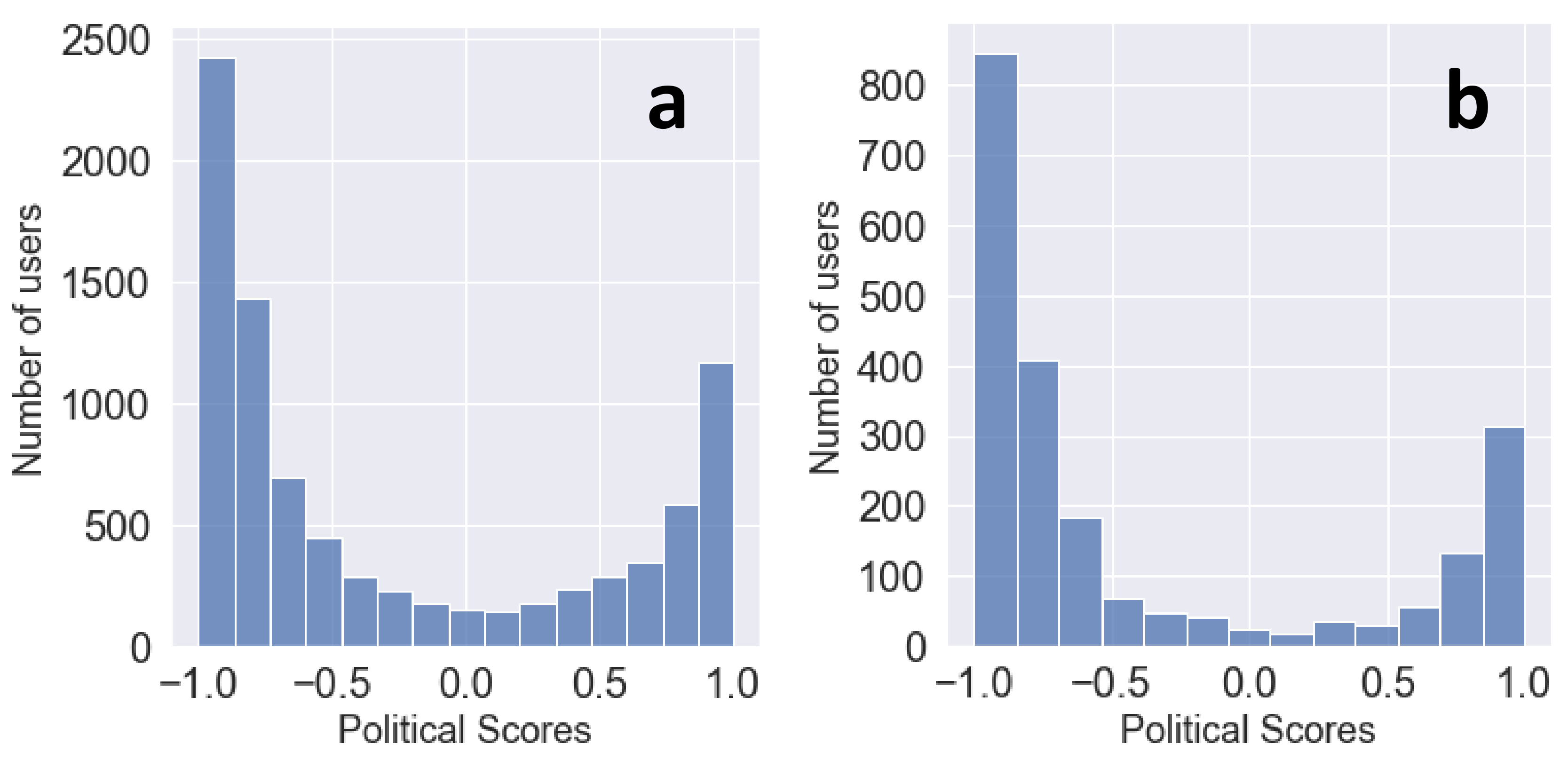}
    % \vspace{-0.4cm}
    \caption{Political scores of (a) the users in the \textit{inflation} dataset and (b) the users in the \textit{vaccine} dataset, inferred by follower relationship on Twitter. Note that the left side of the plot corresponds to the politically left and the right side to the politically right.}
    \label{fig:political}
\end{figure}

The political score has a range of 
\begin{math}
[-1, 1]
\end{math}
with 
\begin{math}
1
\end{math}
meaning the most extreme right and $-1$ meaning the most extreme left. To increase the robustness of inference, only the political scores of the Twitter users who follow at least five accounts from either side (i.e., right- and left-leaning) are calculated. In this way, we calculate the political scores of 8,824 Twitter users in the \textit{inflation} dataset and 2,214 Twitter users in the \textit{vaccine} dataset. Figure~\ref{fig:political} shows the distribution of political scores. More Twitter users are left-leaning, which is \textit{consistent} with the findings of the Pew Research Center, i.e., Twitter users are more likely to identify as Democrats than Republicans.\footnote{\url{https://www.pewresearch.org/internet/2019/04/24/sizing-up-twitter-users/} [Accessed 2022-04-03]}

% \vspace{-0.5cm}

\subsection{Topic Inference}
We apply a Latent Dirichlet Allocation (LDA) model~\cite{blei2003latent} to the tweets of each group to extract the topics. The hyper-parameters are selected based on a grid search and the coherence score. By assigning the dominant topic label to each tweet, we obtain the topic distributions. We refer to the topic with the highest proportion as the most popular topic among the tweets of each group. 

\subsection{Representative Image Selection}
Similar to \citet{chen2017selfie}, we extract a 2048-dimensional feature vector for each image from the last “pool5” layer of ResNet-50. Next, we apply K-means clustering. The optimal number of clusters is chosen based on the Silhouette Coefficient~\cite{rousseeuw1987silhouettes}. Within each cluster, we plot the images that are closest to their corresponding cluster centers in Figure~\ref{fig:user_modal_inflation}c.

% In addition, since the number of images is relatively small, we inspect them manually (72 for {\tt Left}, 177 for {\tt Middle} and 99 for {\tt Right}). In the end, guided by the clustering results, we select the representative figures and plot them in Figure~\ref{fig:user_modal_inflation}c. 

\section{Supplemental Results}
Tables~\ref{tab:user_des_inflation} and \ref{tab:user_des_vaccine} show the word frequency of the  user descriptions of the users in the {\tt Left}, {\tt Middle}, and {\tt Right} clusters. Tables~\ref{tab:topic_inflation} and \ref{tab:topic_vaccine} show the top 15 keywords of the most popular topics of the {\tt Left}, {\tt Middle}, and {\tt Right} clusters.

\begin{table}[htbp]
\scriptsize
\centering
\caption{Word frequency of the user descriptions of the users in the  {\tt Left}, {\tt Middle}, and {\tt Right} clusters in the \textit{inflation} dataset (top 10).}
\begin{tabular}{llllll}
\toprule
\multicolumn{2}{l}{{\tt Left}} & \multicolumn{2}{l}{{\tt Middle}} & \multicolumn{2}{l}{{\tt Right}} \\
word            & count  & word            & count    & word            & count   \\
\midrule
resist       & 228       & fan             & 90       & maga            & 380     \\
blm          & 193       & love            & 80       & conservative    & 261     \\
love         & 181       & former          & 76       & love            & 217     \\
mom          & 150       & writer          & 73       & trump           & 216     \\
lover        & 143       & proud           & 72       & god             & 161     \\
democrat     & 141       & maga            & 68       & patriot         & 146     \\
proud        & 137       & author          & 67       & proud           & 128     \\
writer       & 128       & conservative    & 67       & life            & 127     \\
fan          & 122       & us              & 65       & christian       & 119     \\
retired      & 117       & opinions        & 63       & kag             & 109    \\
\bottomrule
\end{tabular}
\label{tab:user_des_inflation}
\end{table}

\begin{table}[htbp]
\scriptsize
\centering
\caption{Word frequency of the user descriptions of the users in the  {\tt Left}, {\tt Middle}, and {\tt Right} clusters in the \textit{vaccine} dataset (top 10).}
\begin{tabular}{llllll}
\toprule
\multicolumn{2}{l}{{\tt Left}} & \multicolumn{2}{l}{{\tt Middle}} & \multicolumn{2}{l}{{\tt Right}} \\
word            & count  & word            & count    & word            & count   \\
\midrule
resist          & 192    & people          & 14       & maga            & 179     \\
mom             & 92     & blm             & 14       & trump           & 116     \\
blm             & 87     & conservative    & 14       & conservative    & 84      \\
love            & 71     & author          & 13       & love            & 84      \\
lover           & 68     & us              & 12       & kag             & 83      \\
trump           & 67     & maga            & 12       & god             & 68      \\
proud           & 66     & trump           & 11       & patriot         & 54      \\
theresistance   & 64     & mom             & 11       & 2a              & 50      \\
bidenharris     & 63     & love            & 11       & trump2020       & 46      \\
democrat        & 61     & writer          & 11       & family          & 44     \\
\bottomrule
\end{tabular}
\label{tab:user_des_vaccine}
\end{table}

\begin{table}[htbp]
\scriptsize
\centering
\caption{Top 15 keywords of the most popular topics of the {\tt Left}, {\tt Middle}, and {\tt Right} clusters in the \textit{inflation} dataset.}
\begin{tabular}{llllll}
\toprule
\multicolumn{2}{l}{{\tt Left}} & \multicolumn{2}{l}{{\tt Middle}} & \multicolumn{2}{l}{{\tt Right}} \\
word            & weight  & word            & weight    & word            & weight   \\
\midrule
high & 0.011 & high & 0.023 & people & 0.020 \\ 
economic & 0.011 & get & 0.019 & high & 0.016 \\ 
bill & 0.010 & gas & 0.017 & say & 0.015 \\ 
well & 0.009 & year & 0.011 & mean & 0.014 \\ 
economy & 0.008 & rise & 0.011 & gas & 0.013 \\ 
inflationary & 0.008 & border & 0.011 & pay & 0.012 \\ 
get & 0.007 & well & 0.010 & go & 0.011 \\ 
year & 0.007 & pay & 0.009 & economic & 0.010 \\ 
say & 0.007 & covid & 0.008 & work & 0.010 \\ 
president & 0.007 & trump & 0.008 & rise & 0.010 \\ 
people & 0.007 & crime & 0.008 & economy & 0.009 \\ 
much & 0.007 & see & 0.007 & bad & 0.009 \\ 
ease & 0.006 & come & 0.007 & increase & 0.009 \\ 
good & 0.006 & go & 0.007 & crisis & 0.008 \\ 
plan & 0.005 & big & 0.006 & problem & 0.008 \\ 
\bottomrule
\end{tabular}
\label{tab:topic_inflation}
\end{table}

\begin{table}[htbp]
\scriptsize
\centering
\caption{Top 15 keywords of the most popular topics of the {\tt Left}, {\tt Middle}, and {\tt Right} clusters in the \textit{vaccine} dataset.}
\begin{tabular}{llllll}
\toprule
\multicolumn{2}{l}{{\tt Left}} & \multicolumn{2}{l}{{\tt Middle}} & \multicolumn{2}{l}{{\tt Right}} \\
word            & weight  & word            & weight    & word            & weight   \\
\midrule
trump & 0.057 & take & 0.027 & take & 0.029 \\ 
say & 0.015 & get & 0.026 & stop & 0.022 \\ 
get & 0.013 & safe & 0.019 & trump & 0.021 \\ 
debate & 0.013 & people & 0.016 & get & 0.016 \\ 
people & 0.011 & time & 0.014 & say & 0.016 \\ 
know & 0.009 & work & 0.014 & mask & 0.014 \\ 
need & 0.009 & say & 0.014 & kamala & 0.013 \\ 
take & 0.009 & trump & 0.012 & president & 0.012 \\ 
trial & 0.008 & think & 0.012 & work & 0.010 \\ 
safe & 0.008 & believe & 0.012 & testing & 0.008 \\ 
china & 0.008 & effective & 0.012 & go & 0.008 \\ 
make & 0.008 & virus & 0.009 & public & 0.008 \\ 
see & 0.008 & available & 0.009 & administration & 0.008 \\ 
trust & 0.007 & debate & 0.009 & see & 0.008 \\ 
administration & 0.007 & trust & 0.009 & new & 0.007 \\ 
\bottomrule
\end{tabular}
\label{tab:topic_vaccine}
\end{table}

\end{document}